\documentstyle[prl, aps, twocolumn, floats]{revtex}

\begin{document}

\input epsf.sty

\twocolumn[
\hsize\textwidth\columnwidth\hsize\csname@twocolumnfalse\endcsname

\draft

\title{Dissipation-Driven Breakdown of Universality in Two-Dimensional
  Superconductors}

\author{Klaus V\"olker} \address{Department of Physics and Astronomy,
  University of California, Los Angeles, CA 90095} \date{\today}

\maketitle

\begin{abstract}
  The influence of gapless dissipative degrees of freedom on the
  superconductor-insulator transition in two dimensions is
  investigated. We develop a series expansion for the free energy of a
  (2+1)-dimensional XY model coupled to a bosonic heat bath that can
  be approximately summed to all orders. The calculation
  explicitly conserves topological excitations. We derive the 
  zero temperature phase diagram and
  the free energy critical exponent, and find a
  transition from universal to non-universal scaling behavior as the
  coupling to the dissipative environment is increased, implying the 
  existence of a new universality class.
\end{abstract}


]

\newcommand{\EJ}{E_{J}} 
\newcommand{\EC}{E_{C}} 
\newcommand{\ECI}{\EC^{-1}}
\newcommand{\nn}{ \nonumber \\ }

\newcommand{\inversealpha}{ \gamma }

The subject of
universality in the zero-temperature 
superconductor-insulator transition in two
dimensions has been a debated issue for some time.
Early experiments on thin superconducting films\cite{Orr} seemed to
indicate that the critical resistance right at the transition is
always in the vicinity of the resistance quantum \(
R_{Q}=h/(2e)^{2}\simeq 6.5k\Omega \).  This universality could be
explained by a scaling analysis of the bosonic Hubbard
model\cite{Fisher44}, which can be approximately mapped to the XY
Hamiltonian, Eq.~(\ref{eqn_H0}). In later experiments very different
critical resistivities were measured in amorphous
films\cite{YazKapitulnikMarkovich} 
and Josephson junction arrays\cite{Geerligs},
 so that the universality hypothesis was cast into 
doubt.
%Furthermore, in experiments in an applied magnetic field a 
% ``metallic''
%regime\cite{YazKapitulnik,Metallic} with finite zero-temperature 
%conductivity has been found.

The universality class of the superconductor-insulator transition
 is affected by disorder and by
coupling to dissipative degrees of freedom
of the environment. In this paper we will concentrate 
on the latter,
leaving the disordered model for future study.
In a Josephson array with
zero or weak dissipation
 the phase transition is driven by a competition between
the Josephson coupling energy
and the charging energy and (at $T=0$) belongs to the (2+1)d XY 
universality class. It is well known, on the other hand, that even in a 
single Josephson
junction the order parameter phase can localize  
if the electronic degrees
of freedom are taken into account, and if the associated density of
states extends down to zero frequency\cite{Schmid}. 
Clearly, under the same
conditions, an array of such Josephson junctions would display phase
order, and hence superconductivity, as well. Then the transition
would have an essentially  
 (0+1)-dimensional character, thereby motivating the
existence of a new universality class.

\begin{figure}
  \centerline{\epsfxsize=3.4in \epsffile{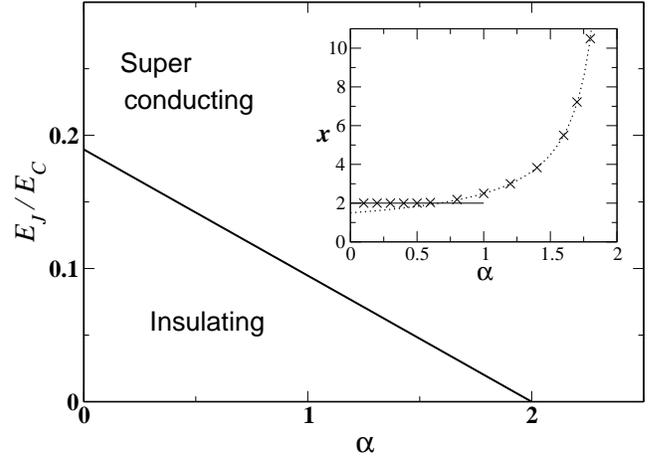}}
  \caption{The zero-temperature
phase diagram in the non-intersecting path
approximation. 
%in combination with the conjectured renormalization flow diagram. 
%The isolated star marks the critical point of the non-dissipative XY transition, the other stars label a critical line. The diamonds mark a line of stable fixpoints at $\EJ/\EC=0$.
The X's in the inset show the free energy 
critical exponent as calculated here. 
The solid and dotted lines in the inset show
the functional form given in Eq.~\ref{eqn_exponent}. }
\label{fig_phasediagram}
\end{figure}

\begin{figure}
  \centerline{\epsfxsize=3.1in \epsffile{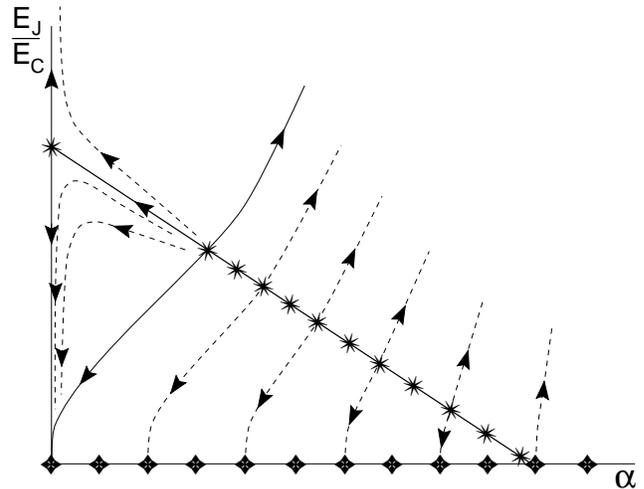}}
  \caption{The conjectured renormalization group flow diagram. The isolated star marks the critical point of the non-dissipative XY transition, the other stars label a critical line. The diamonds mark a line of stable fixpoints at $\EJ/\EC=0$.}
  \label{fig_rgflow}
\end{figure}

In this letter we will develop a framework in which 
these universality classes can be understood in a rigorous,
quantitative way. 
Our formalism, which constitutes an expansion
about the insulating state, explicitly preserves the $2\pi$ periodicity
of the phase variables, and hence the topological excitations (vortices), 
which are known to drive the phase transition in the classical limit.
We  derive a series expansion for the free
energy of the associated (2+1)-dimensional statistical mechanics model, 
which can be approximately summed to all orders.
At $T=0$, this is of course just the ground state energy of
the quantum system.
Strikingly, we find a crossover from universal to non-universal
scaling behavior as the coupling strength $\alpha$
to the environment
is increased, as shown in the inset of Fig.~(\ref{fig_phasediagram}): 
As long as $\alpha < 2/3$, the free 
energy critical exponent is unchanged from its value for the
non-dissipative model, while it  varies 
continuously  for $\alpha > 2/3$, implying  a new 
universality class. Fig.~(\ref{fig_rgflow}) shows a conjectured 
renormalization group (RG)
flow diagram which we will comment on
at the end of the paper.

The model under consideration has been
investigated in the past, often by approximations that, in one way or
the other, amount to some variant of a self-consistent field
approximation, or to linearizing the Josephson coupling term, and hence
destroying the \( 2\pi \) periodicity. However, a recent investigation
\cite{Wagenblast} utilizing a Ginzburg-Landau-Wilson 
formulation showed results very similar to ours.

A number of mechanisms (see e.g.\onlinecite{Wagenblast} for
references)  can give rise to an ungapped
electronic density of states, and hence Ohmic damping, in thin-film
superconductors. As a scenario
that recently gained experimental support\cite{Chervenak} we mention
the formation of local pools of unpaired electrons due to spatial
fluctuations of the order parameter amplitude, caused by impurities. 
Other possibilities include d-wave high-temperature superconductors 
with nodal order parameter, coupling to electronic
degrees of freedom in the substrate and in the vortex cores, or Andreev
scattering. 
A Josephson array with 
tunable coupling to the
environment in the form of a 2d electron gas has actually been
constructed by Clarke {\it et al}.\cite{Clarke}

Two-dimensional Josephson junction arrays, as well as superconducting
thin films can be modeled as a system of coupled quantum rotors
with the Hamiltonian (in the absence of dissipation)

\begin{equation}
\label{eqn_H0}
H_{XY} = \frac{\EC}{2} \sum_{x} n_{x}^{2} + \EJ 
\sum_{\left\langle xy\right\rangle} 
\left[ 1-\cos (\phi_{x}-\phi_{y})\right],
\end{equation}
where the phase \( \phi \) of the superconducting order parameter
is a compact variable on the interval \(
[0,2\pi ] \), and the Cooper pair number operator \( n \) is its
canonical conjugate. $\EC=(2e)^2/C$ and $\EJ$ are the charging energy
and the Josephson coupling energy, respectively. In the
above Hamiltonian the ground state is limited to integer filling factors,
and more
generally we would have to replace \( n_{x}^{2} \) by \( (n_{x}-\delta
)^{2} \), where \( \delta \) is some non-integer number. This difference
is critical in the
 non-dissipative model \cite{Fisher44}, 
where a non-integer filling factor increases
charge fluctuations and hence enhances phase ordering. 
However, we will see below that this dependence on the filling factor 
disappears as soon as we include Ohmic dissipation.

In an imaginary time path integral formalism the action for
the dissipative system reads

\begin{eqnarray}
S[\phi ] & = & \int_0^\beta d\tau \left[ 
\frac{1}{2\EC} \sum_x \dot{\phi }_{x\tau }^{2}-
\EJ \sum_{\left< x y \right>}
\cos (\phi_{x\tau} - \phi_{y\tau})\right] \nn 
 & + & \frac{1}{2}\sum _{x}\int_0^\beta d\tau \int_0^\beta d{\tau'}
 \phi _{x\tau }D(\tau -\tau ')\phi _{x\tau '},\label{eqn_dissipativeaction} 
\end{eqnarray}
where the last term arises from an
integration over the dissipative degrees of freedom
\cite{Sudip,CaldeiraLeggett} and represents Cooper pair breaking
processes due to the interaction with the environment. An alternative
model would couple the dissipative degrees of freedom to 
the {\em phase difference} across the junction, reflecting the
existence of a normal conducting channel between the islands, but 
will not be considered here. 
For Ohmic damping the kernel \( D(\tau ) \) is given
by its Fourier transform

\begin{equation}
\label{eqn_dissipativekernel}
D(\omega ) = \frac{\alpha }{2\pi }|\omega| 
% \quad \mbox {for}\: |\omega |\ll \omega _{c}.
\end{equation}
for $|\omega|$ less than a high-energy cutoff $\omega_c$, which can be
approximately set equal to some physical cutoff, such as the 
bandwidth of localized electron pools.
The dimensionless parameter $\alpha$ controls the strength of the 
dissipative coupling.

In general the compactness of the
angular variables causes \( \phi \) and \( \phi +2\pi \)
to be identified, so that \( \phi _{\tau } \) is periodic only up to an
integer multiple of \( 2\pi \). We therefore define \( \tilde{\phi
  }_{x\tau }=\phi _{x\tau }-2\pi \beta ^{-1}m_{x}\tau _{,} \) where \(
\tilde{\phi }_{x\tau } \) satisfy periodic boundary conditions and the
\( m_{x} \) are winding numbers. 
An infrared divergence in the dissipative term of
(\ref{eqn_dissipativeaction}), however, 
effectively suppresses all nonzero values of \( m_{x} \).
%%%%%% \cite{ref_InfraredDivergence}.  
This leads to another simplification as well: The
non-integer filling factors mentioned above give rise to a term \(
i\delta \int _{0}^{\beta }d\tau \dot{\phi }_{x\tau }=2\pi i\delta
m_{x} \) in the action. Hence, as long as winding numbers are
suppressed, non-integer filling factors are of no consequence.
Intuitively speaking, since in our model the number of Cooper pairs is
not conserved, any dependence on the filling factor is washed out.

We now express the partition function
as a power series in $\EJ$:

\begin{eqnarray}
Z & = & \sum _{n=0}^{\infty }\frac{1}{n!}\prod ^{n}_{i=1}\left[ \frac{\EJ }{2}\int_0^\beta d\tau _{i}\sum _{[x_{i}y_{i}]}\right] \: \int _{0}^{2\pi }D\phi _{x\tau } e^{-S},\nn \nonumber \\
S & = & \exp \left\{ -S_{0}+i\sum _{x\tau }\int d\tau \eta _{x\tau }\phi _{x\tau }\right\}. \label{eqn_PartitionFunction} 
\end{eqnarray}

Here \( [x_{i}y_{i}] \) indicates a \emph{directed} bond between
two neighboring lattice sites, and the sum runs over all such 
directed bonds.
\( \eta _{x\tau }=\sum _{i=1}^{n}\delta (\tau -\tau_{i})
(\delta _{x,x_{i}}-\delta _{x,y_{i}}) \)
can be interpreted as a ``charge density,'' with $x_i$ and $y_i$ 
indicating the position of positive and negative charges,
respectively. Each directed pair \( [x_{i}y_{i}] \) then 
constitutes a ``dipole.''
\( \EJ /2 \) plays the role of a fugacity, and
\( S_{0} \) is the action (\ref{eqn_dissipativeaction}) with \( \EJ =0 \).
We may call the resulting model the Coulomb
dipole model, in analogy to the Coulomb gas model for the resistively
shunted Josephson junction\cite{Schmid}. 
After integrating out the phase variables the action reads
\begin{equation}
S=\frac{1}{2}\sum _{x}\int_0^\beta d\tau \int_0^\beta d\tau '\eta _{x\tau }v(\tau
-\tau ')\eta _{x\tau '},
\label{eqn_DipoleAction}
\end{equation} with

\[
v(\tau )=\frac{1}{\beta }\sum _{\omega }\frac{e^{i\omega \tau
    }-1}{\ECI \omega ^{2}+D(\omega )},\]
where
we have subtracted out an infinite constant that
enforces ``charge neutrality'' for each lattice site. For the rest of
the paper we will assume \( \EC \ll \omega _{c} \). 
Then the interaction potential
is well approximated by

\begin{equation}
v(\tau )\simeq -\inversealpha \ln \left( \frac{\beta \omega
    ^{*}}{\pi }\sin \frac{\pi |\tau |}{\beta }+1\right) ,
\label{eqn_Interaction}
\end{equation} where
$\inversealpha = 2/\alpha$.
This amounts to replacing the kinetic energy term \( \ECI \omega ^{2} \) by
an effective cutoff 
%%%%%%%%%%%%%%%%%%%%%%%%%%%%%%%%%%%%%%%%%%%%%%%%%%%%%%%%%%%%%%%%%%%
\( \omega ^{*} = \EC / 2\inversealpha \).
%%%%%%%%%%%%%%%%%%%%%%%%%%%%%%%%%%%%%%%%%%%%%%%%%%%%%%%%%%%%%%%%%%%
% \( \omega ^{*} \simeq 0.567 \EC / \inversealpha \).
%We can view each term in the sum (\ref{eqn_PartitionFunction}) 
%as a collection of dipoles that are
%oriented along the x- or y-axis and interact with a long-range
%potential in the imaginary time direction. 
%The partition sum can now be viewed as a sum over all neutral 
%configurations of dipoles, which are oriented along the x- or y-axis
%and interact with a logarithmic potential in the imaginary-time 
%direction. 
Each term in (\ref{eqn_PartitionFunction}) can be
represented by diagrams such as those in Eq.~(\ref{eqn_diagrams}). 
Here each arrow corresponds to
a ``dipole'' as defined above, or in physical terms to a Cooper pair
transfer event. Double arrows, such as those in $D_2$,  indicate two 
dipoles of opposite orientation occupying the same bond. 
Similarly, the first diagram in $D_{4c}$
has four dipoles occupying the same bond. As in standard diagrammatic
perturbation theory we can use the linked cluster theorem to write the
perturbation series for the free energy as a sum over connected
diagrams.  The diagrams for the first few terms in the expansion

\begin{eqnarray*}
f=F/N & = & f_{0}-2\left( \frac{\EJ }{2}\right) ^{2}D_{2}\\
 & - & \left( \frac{\EJ }{2}\right) ^{4}
 \left\{ 2D_{4a}+6D_{4b}+\frac{1}{2}D_{4c}\right\} -\cdots 
\end{eqnarray*}
 follow directly from 
Eqs.~(\ref{eqn_PartitionFunction}) to (\ref{eqn_Interaction}):

\newcommand{\vertex}{\circle*{0.2}}
\setlength{\unitlength}{1cm}

\begin{picture}(7,1.5)
\thicklines

\put(0,0.6){ \parbox{1cm}{ \[ D_2 = \] } }
\put(1.5,0.7){\vertex}
\put(1.9,0.7){\vector(1,0){0.3}}
\put(1.9,0.7){\vector(-1,0){0.3}}
\put(2.3,0.7){\vertex}
\put(2.5,0.6){ \parbox{4cm}{ \[
    = \frac{1}{\beta}\int _{0}^{\beta }d\tau _{1}
    \int _{0}^{\beta }d\tau _{2}\frac{1}{f^{2}(\tau _{1}-\tau _{2})},
\] } }

\end{picture}

\begin{picture}(7,1.5)
\thicklines

\put(0,0.8){ \parbox{1cm}{ \[ D_{4a} = \] } }
\put(1.5,1.3){\vertex}
\put(1.5,0.5){\vertex}
\put(2.3,1.3){\vertex}
\put(2.3,0.5){\vertex}
\put(2.2,1.3){\vector(-1,0){0.4}}
\put(1.8,1.3){\line(-1,0){0.2}}
\put(1.6,0.5){\vector(1,0){0.4}}
\put(2.0,0.5){\line(1,0){0.2}}
\put(1.5,1.2){\vector(0,-1){0.4}}
\put(1.5,0.8){\line(0,-1){0.2}}
\put(2.3,0.6){\vector(0,1){0.4}}
\put(2.3,1.0){\line(0,1){0.2}}
\put(2.5,0.8){ \parbox{3cm}{ \[
    = \frac{1}{\beta}\int _{0}^{\beta }\frac{d\tau _{1}\cdots d\tau _{4}}
    {f_{12}f_{23}f_{34}f_{41}},
\] } }

\end{picture}

\begin{picture}(7,1.6)
\thicklines

\put(0,1.4){ \parbox{1cm}{ \[ D_{4b} = \] } }
\put(1.5,1.5){\vertex}
\put(1.9,1.5){\vector(1,0){0.3}}
\put(1.9,1.5){\vector(-1,0){0.3}}
\put(2.3,1.5){\vertex}
\put(2.7,1.5){\vector(1,0){0.3}}
\put(2.7,1.5){\vector(-1,0){0.3}}
\put(3.1,1.5){\vertex}

\put(1.8,1.4){ \parbox{5cm}{ \[ - \; ( \quad \quad \quad \; )^2 \] } } 

\put(4.1,1.5){\vertex}
\put(4.5,1.5){\vector(1,0){0.3}}
\put(4.5,1.5){\vector(-1,0){0.3}}
\put(4.9,1.5){\vertex}

\put(0.0,0.6){ \parbox{8cm}{ \[
    = \frac{1}{\beta}\int _{0}^{\beta }\frac{d\tau _{1}\cdots d\tau _{4}}
    {f^{2}_{23}f^{2}_{41}}\left\{ \frac{f_{13}f_{24}}{f_{12}f_{34}}-1\right\},
\] } }

\end{picture}

\begin{picture}(7,1.7)
\thicklines

\put(0,1.4){ \parbox{1cm}{ \[ D_{4c} = \] } }

\put(1.5,1.5){\vertex}
\put(1.9,1.5){\vector(1,0){0.25}}
\put(1.9,1.5){\vector(-1,0){0.25}}
\put(2.15,1.5){\vector(1,0){0.15}}
\put(1.65,1.5){\vector(-1,0){0.15}}
\put(2.3,1.5){\vertex}

\put(1.0,1.4){ \parbox{5cm}{ \[ - \; 2 ( \quad \quad \quad \; )^2 \] } } 

\put(3.4,1.5){\vertex}
\put(3.8,1.5){\vector(1,0){0.3}}
\put(3.8,1.5){\vector(-1,0){0.3}}
\put(4.2,1.5){\vertex}

\put(0.0,0.6){ \parbox{8cm}{ 
\begin{equation}
  = \frac{1}{\beta}\int _{0}^{\beta }\frac{d\tau _{1}\cdots d\tau _{4}}
  {f^{2}_{23}f^{2}_{41}}\left\{ \frac{f^{2}_{13}f^{2}_{24}}{f^{2}_{12}
      f^{2}_{34}}-2\right\}. 
  \label{eqn_diagrams}
\end{equation} 
} }

\end{picture}

%\begin{eqnarray*}
%D_{2} & = & \frac{1}{\beta}\int _{0}^{\beta }d\tau _{1}\int _{0}^{\beta }d\tau _{2}\frac{1}{f^{2}(\tau _{1}-\tau _{2})}\\
%D_{4a} & = & \frac{1}{\beta}\int _{0}^{\beta }\frac{d\tau _{1}\cdots d\tau _{4}}{f_{12}f_{23}f_{34}f_{41}}\\
%D_{4b} & = & \frac{1}{\beta}\int _{0}^{\beta }\frac{d\tau _{1}\cdots d\tau _{4}}{f^{2}_{23}f^{2}_{41}}\left\{ \frac{f_{13}f_{24}}{f_{12}f_{34}}-1\right\} \\
%D_{4c} & = & \frac{1}{\beta}\int _{0}^{\beta }\frac{d\tau _{1}\cdots d\tau _{4}}{f^{2}_{23}f^{2}_{41}}\left\{ \frac{f^{2}_{13}f^{2}_{24}}{f^{2}_{12}f^{2}_{34}}-2\right\} .
%\end{eqnarray*}

Here
\[
f(\tau )=e^{-v(\tau )}=\left( \frac{\beta \omega ^{*}}{\pi }\sin
  \left| \frac{\pi \tau }{\beta }\right| +1\right) ^\inversealpha,\] and
\( f_{ij} \) is a shorthand for \( f(\tau _{i}-\tau _{j}) \). 
The subtractions in \( D_{4b} \) and \( D_{4c} \) are
corrections for double counting and are distributed such as
to render the diagrams finite in the zero-temperature limit, for $\alpha < 2$. 
Beyond this value
each diagram is divergent at $T=0$,
reflecting the phase localization transition in a single junction
mentioned above.

While for \( n\ge 4 \) the diagrams are generally
very hard to evaluate, there exists a subset which
can be computed exactly (within numerical limits) to all orders. 
The following results are based on this partial
summation, and we will comment on the validity of this approximation
below. The diagrams we consider are the non-intersecting loops 
such as \( D_{4a} \)  in Eq.~(\ref{eqn_diagrams}). 
We then have

\[
D_{na}  =  \frac{1}{\beta}\int _{0}^{\beta }\frac{d\tau _{1}\cdots d\tau _{n}}{f_{12}f_{23}\cdots f_{n1}}
  =  \int _{-\infty }^{\infty } \frac{d\omega}{2\pi} \, g^{n}(\omega ),
\]
where (at zero temperature)

\begin{eqnarray*}
g(\omega ) & = & \int _{-\infty }^{\infty }d\tau \frac{e^{i\omega \tau }}{\left[ |\omega ^{*}\tau |+1\right] ^\inversealpha}\\
 & = & \frac{(\omega ^{*})^{-\inversealpha }}{(i\omega )^{1-\inversealpha }}
 \Gamma \left( 1-\inversealpha,i\omega \right) +c.c.
\end{eqnarray*}

The  numerical prefactor is equal to the number of translationally
distinct self-avoiding polygons of length \( n \) on a square lattice,
which is known\cite{RandomWalks} to behave
asymptotically as \( A_{2n}\sim \mu ^{2n} \), with the connective
constant \( \mu \simeq 2.65 \). 
Hence the free energy is, in this approximation,
\begin{equation}
\label{eqn_RPAfreeenergy}
f=f_{0}-\frac{\omega ^{*}}{2\pi} 
\sum ^{\infty }_{n=1}A_{2n}
\left( \frac{\EJ}{2\omega ^{*}}\right) ^{2n}G_{2n},
\end{equation}
where \( G_{2n}=(\omega ^{*})^{2n-1}\int _{-\infty }^{\infty }d\omega \,
g^{2n}(\omega ) \) can easily be evaluated numerically. This
 series constitutes an expansion about a disordered
fixpoint of the RG flow diagram.  The free energy
is singular at the phase transition, and according to the fundamental
properties of analytic functions this non-analyticity determines the
radius of convergence. We can therefore derive the phase boundary from the
asymptotic behavior of the coefficients in (\ref{eqn_RPAfreeenergy}):
\[
\left( \frac{\EJ }{2\omega ^{*}}\right) _{c}=\lim _{n\to \infty }\left[
  \mu^{2n}G_{2n}\right] ^{-1/2n}.\]

Numerically we find that
$ G_{n}^{-1/n}\to (\inversealpha-1)/2$ for large \( n \). 
This relation holds to an accuracy of more than
five digits, and is presumably exact.
% so that the phase boundary is
% given by \( (\EJ /\omega ^{*})_{c}=(1-\alpha )/(\mu \alpha ) \).
Recalling 
%%%%%%%%%%%%%%%%%%%%%%%%%%%%%%%%%%%%%%%%%%%%%%%%%%%%%%%%%%%%%%%%%%%
 \( \omega ^{*} = \EC / 2\inversealpha \)
%%%%%%%%%%%%%%%%%%%%%%%%%%%%%%%%%%%%%%%%%%%%%%%%%%%%%%%%%%%%%%%%%%%
% \( \omega ^{*} \simeq 0.567 \EC / \inversealpha \)
 and $\inversealpha = 2/\alpha$ we
find that the phase
boundary is at
\begin{equation}
\label{eqn_phaseboundary}
%%%%%%%%%%%%%%%%%%%%%%%%%%%%%%%%%%%%%%%%%%%%%%%%%%%%%%%%%%%%%%%
 \left( \frac{\EJ }{\EC }\right) _{c}=\frac{2-\alpha }{4\mu }.
%%%%%%%%%%%%%%%%%%%%%%%%%%%%%%%%%%%%%%%%%%%%%%%%%%%%%%%%%%%%%%%
% \left( \frac{\EJ }{\EC }\right) _{c}=\frac{2-\alpha }{7.92 }.
\end{equation}

The phase diagram is shown in Fig.~(\ref{fig_phasediagram}). 
In order to assess the quality of the non-intersecting
path approximation it is important to realize that we did not
omit any terms in the  series expansion, but merely
 replaced the more
complicated diagrams with products of simpler ones. For example, the
connected piece of diagram \( D_{4b} \) is replaced by \( (D_{2})^{2} \). 
To estimate the errors introduced by this approximation we
calculated all intersecting diagrams 
up to sixth order by Monte Carlo integration\cite{me}.
We find that in fourth order these corrections change the 
value of $G_n^{1/n}$ by a factor between $1.1$ 
(for $\alpha \to 0$) and  $1.25$ (at $\alpha=2$). 
For the 
sixth-order term these factors are $1.12$ ($\alpha \to 0$)
and $1.38$ ($\alpha=2$). The second-order term is exact.
Because of the relatively small ratios we
expect our approximation to overestimate the critical value of 
$\EJ/\EC$ somewhat, but not to alter the qualitative features of the
phase diagram. 
In particular, it cannot affect the universality classes.  
%In particular, it cannot
%change the existence of two universality classes.

We now turn to the  free energy critical exponent. Near the phase
transition the singular part of the free energy 
must be of the form \(
(1-y)^{x} \), up to logarithmic corrections,
where \( y=(\EJ/{\EJ}_{0})^{2} \) and $x$ is some 
unknown exponent. Expanding this expression in terms of \( y \) we
get $f \sim \sum_n c_{2n} y^n$ with

\[
c_{2n} = \frac{(-1)^{n}\Gamma (x +1)}{\Gamma (n+1)\Gamma (x
  -n+1)}.\] 

The leading
asymptotic behavior of the coefficients is
given by \( \ln c_{2n}\sim -(x+1)\ln n \). The
corresponding terms in (\ref{eqn_RPAfreeenergy}) behave as

\begin{eqnarray*}
  \ln A_{2n} &\sim& 2n \ln \mu + (\alpha_s - 3)\ln n, \\
  \ln G_{2n} &\sim& 
  2n \ln \frac{\inversealpha - 1}{2}
  -\max \left( \frac{1}{2} , \;  \frac{1}{\inversealpha-1} \right)\ln n,
\end{eqnarray*}
where the first relation can be found in the mathematical 
literature\cite{RandomWalks}
(the exponent $\alpha_s$ is conjectured to be $1/2$), and the second relation
has been determined numerically. We expect the numerical estimates 
to be correct at least to three digits precision. 
Combining these terms we find that the free 
energy critical exponent is

\begin{equation}
x = \max \left( 2, \; \frac{1}{2} + \frac{2}{2-\alpha} \right),
\label{eqn_exponent}
\end{equation}
which is shown in the inset of Fig.~(\ref{fig_phasediagram}). Due to the high
quality of the fits shown we assume that this is an exact result with only 
minor corrections close to $\alpha=2/3$. 
%The striking feature here is the apparent universality 
%of \( x \) for weak
%coupling to the environment, and the breakdown of universality for
%larger coupling $\alpha > 1/3$. A similar breakdown of universality 
%has been found
The consistency between our results and those of Ref.\onlinecite{Wagenblast}
is emphasized by the scaling relation $z=2x-3$ between $x$ and the dynamical exponent $z$ 
calculated there. 
%We want to emphasize the similarity between the functional form of $x$
%and that of the dynamic exponent $z$ as calculated in  
%Ref.\onlinecite{Wagenblast}. Incidentially, the two quantities satisfy
%the scaling relation $z=2x-3$ for all $\alpha < 1$. 
%in the critical conductance by the authors of 
%Ref.\onlinecite{Wagenblast},
%who used very different methods. Incidentially, their result for the 
%dynamical exponent \( z \) yields the scaling relation $z=2x-3$, 
%which is valid for all $\alpha < 1$,
%when compared with our result for $x$.  
Since all correlation functions and
thermodynamic properties of the system, as well as transport
properties, can be derived from the free energy, this transition from
universal to non-universal scaling behavior should be observable in
those quantities as well.

In the language of the renormalization group, 
the critical point which controls
the phase transition of the non-dissipative system has a finite basin
of attraction.  Since
the flow equations are presumably analytic functions of the coupling
parameters this implies the existence of an additional critical point
on the phase boundary, at \( \alpha =2/3 \). The continuously varying
critical exponent furthermore suggests a line of critical points
beyond that value. 
The first-order  RG equations \cite{Sudip} 
%\begin{eqnarray*}
%d/dl \: \ln(\EJ/\EC) & = & -2/\alpha, \;\; \mbox{and}\\ 
%d/dl \: \ln \alpha & = & 0,
%\end{eqnarray*}
$d\EJ/dl = (1-2/\alpha)\EJ$ and 
$d\alpha/dl = 0$
($l$ is a scale factor for the high-energy cutoff)
determine the lower portion of the RG flow diagram,
and in particular imply that \( \EJ /\EC =0 \) constitutes a
stable fixed line. 
Together these arguments lead to the  RG flow diagram 
shown in Fig.~(\ref{fig_rgflow}). 

I would like to thank Sudip Chakravarty for enlightening discussions,
and Chetan Nayak for useful comments.
This work was supported by the grant NSF-DMR-9971138.

\end{document}